%% file: 9410379.tex
\documentstyle{ichep}
\input{matex}
\begin{document}
\newcommand{\dis}{\displaystyle}
\title{{\LARGE{\bf Beyond the Standard Model in the Lepton Sector}}}
\author{ {\large {\bf Jos\'e W. F. Valle $^{\dag}$ } }}
\address{\dag\
Departament de F\'{\i}sica Te\'{o}rica, Universitat de Val\`{e}ncia
and \\ Instituto de F\'{\i}sica Corpuscular - C.S.I.C.,
E-46100 Burjassot, Val\`encia, SPAIN         }
\abstract
{I review some of the physics motivations and potential of
various extensions of the \sm that pertain to the lepton sector.
These include extensions of the lepton multiplet content,
closely related to the properties of neutrinos, extensions
of the electroweak breaking sector, such as supersymmetry,
as well as possible extensions of the gauge sector. They may
all lead to new signatures at levels accessible to experiment.}

\twocolumn[\maketitle]

\section{Introduction.}
\hspace{0.5cm}

Our present standard model leaves open many
of the fundamental issues in particle physics,
such as the mechanism of mass generation and the
properties of neutrinos. Extensions of the basic
picture that seek to address these issues, such
as higher unification and supersymmetry, may
lead to extensions of the lepton multiplet
and/or Higgs boson content, and thereby affect
the physics of the lepton sector in an important
way that fortunately can be probed in a variety of
present and future experiments.

\section{Neutral Heavy Leptons.}
\hspace{0.5cm}

There are many motivations to extend the lepton sector
of the electroweak theory. Extra heavy leptons may arise
in models with a higher unification, for example those
with left-right symmetry \cite{LR} or superstrings
\cite{SST}. These models may contain isosinglet neutral
heavy leptons and typically, also neutrino masses \cite{fae}.

They may induce lepton flavour violating (LFV) decays \sa
$\mu \rightarrow e \gamma$, which are exactly forbidden
in the standard model. Although these are a generic
feature of models with massive \neus, in some cases,
they may proceed in models where \neus are strictly
massless \cite{fae}-\cite{CP}.

In the simplest models of seesaw type \cite{GRS} the NHLS are
superheavy so that the expected rate for LFV processes
is expected to be low, due to limits on \neu masses.
However, in other variants \cite{SST} this is not the case
\cite{BER,CP} and this suppression need not be present.
Indeed, present constraints on weak universality violation
allow for decay branching ratios larger than the present
experimental limits \cite{3E} so that these already
are probing the masses and admixtures of the
NHLS with considerable sensitivity. Similar estimates
can be done for the corresponding tau decays \cite{3E,Pila}.
The results are summarized in table 1. See also
figures 5 and 6 given in ref. \cite{Pila}.
Clearly these branching ratios lie within the
sensitivities of the planned tau and B factories,
as shown in ref. \cite{TTTAU}.
\begin{table}
\begin{center}
\caption{Allowed $\tau$ decay branching ratios
}.
\begin{displaymath}
\begin{array}{|c|cr|}
\hline
\mbox{channel} & \mbox{strength} & \mbox{} \\
\hline
\tau \rightarrow e \gamma ,\mu \gamma &  \lsim 10^{-6} & \\
\tau \rightarrow e \pi^0 ,\mu \pi^0 &  \lsim 10^{-6} & \\
\tau \rightarrow e \eta^0 ,\mu \eta^0 &  \lsim 10^{-6} - 10^{-7} & \\
\tau \rightarrow 3e , 3 \mu , \mu \mu e, \etc &  \lsim 10^{-6} - 10^{-7} & \\
\hline
\end{array}
\end{displaymath}
\end{center}
\end{table}

The physics of rare $Z$ decays nicely complements what
can be learned from the study of rare LFV muon and tau decays.
The stringent limits on $\mu \rightarrow e \gamma$ preclude any
possible detectability at LEP of the corresponding
$Z \rightarrow e \mu$ decay. While experimentally closer,
under realistic luminosity and experimental resolution
assumptions, it is still unlikely that one will be able
to see even the $e\tau$ or $\mu\tau$ decays of the Z at
LEP \cite{ETAU}. In any case, there have been dedicated
searches which have set good limits \cite{opal}.

If lighter than the $Z$, NHLS may also be produced in
Z decays such as
\footnote{There may also be CP violation in lepton
sector, even when the known \neus are strictly massless
and lead to Z decay asymmetries \O($10^{-7}$) \cite{CP}} \cite{CERN},
\begin{equation}
Z \rightarrow N_{\tau} + \nu_{\tau}
\end{equation}
Note that the isosinglet neutral heavy lepton
\Nt is singly produced, through the off-diagonal
neutral currents characteristic of models containing
doublet and singlet leptons \cite{2227}.
Subsequent \Nt decays would then give rise to
large missing energy events, called zen-events.
As seen in table 2 this branching ratio can be
as large as $\lsim 10^{-3}$ a value that is already
superseded by the good limits on such decays from
the searches for acoplanar jets and lepton pairs from $Z$
decays at LEP, although some inconclusive hints have
been recently reported by ALEPH \cite{opal}
\begin{table}
\begin{center}
\caption{Allowed branching ratios for rare $Z$
decays. }
\begin{displaymath}
\begin{array}{|c|cr|}
\hline
\mbox{channel} & \mbox{strength} & \mbox{} \\
\hline
Z \rightarrow \Nt \nt &  \lsim 10^{-3} & \\
Z \rightarrow e \tau &  \lsim 10^{-6} - 10^{-7} & \\
Z \rightarrow \mu \tau &  \lsim 10^{-7} & \\
\hline
\end{array}
\end{displaymath}
\end{center}
\end{table}

Finally we note that there can also be large rates for
lepton flavour violating decays in models with radiative
mass generation \cite{zee.Babu88}. For example, this is
the case in the models proposed to reconcile present
hints for \neu masses \cite{DARK92}. The expected decay
rates may easily lie within the present experimental
sensitivities and the situation should improve at PSI
or at the proposed tau-charm factories.

\section{Supersymmetry.}
\hspace{0.5cm}

If supersymmetry exists at the TeV scale it helps
to stabilize the gauge hierarchy problem, one of the
central issues in particle theory today. The most
conventional realization of the idea of supersymmetry
postulates the conservation of R parity.
As a result of this {\sl ad hoc} selection rule, in
the so-called minimal supersymmetric standard model
SUSY particles are only produced in pairs, with the
lightest of them (LSP) being stable.

Nobody knows the origin of this R parity symmetry
and why it is there. There are many ways to break
it, either explicitly or spontaneously (RPSUSY
models).
If R parity is broken spontaneously it shows up
primarily in the couplings of the W and the Z,
leading to rare $Z$ decays such as the single
production of the charginos and neutralinos
\cite{ROMA}, for example,
\begin{equation}
Z \rightarrow \chi^{\pm} \tau^{\mp}
\end{equation}
where the lightest chargino mass is assumed to be smaller than
the Z mass. In the simplest models, the magnitude of R parity
violation is correlated with the nonzero value of the \nt mass
and is restricted by a variety of experiments. Nevertheless
the R parity violating Z decay branching ratios, as an example,
can easily exceed $10^{-5}$, well within present LEP sensitivities.
Similarly, the lightest neutralino (LSP) could also be
singly-produced as $Z \rightarrow \chi^0 \nu_\tau$ \cite{ROMA}.
Being unstable due to R parity violation, $\chi^0$ is
not necessarily an origin of events with missing energy,
since some of its decays are into charged particles.
Thus the decay $Z \rightarrow \chi^0 \nu_\tau$ would give rise to
zen events, similar to those of the MSSM but where the missing
energy is carried by the \nt. Another possibility for zen events
in RPSUSY is the usual pair neutralino production process, where
one $\chi^0$ decays visibly and the other invisibly. The
corresponding zen-event rates can be larger than in the MSSM.

Although the \nt can be quite massive in these models,
it is perfectly consistent with cosmology \cite{KT}
including primordial nucleosynthesis \cite{BBNUTAU},
since it decays sufficiently fast by majoron emission
\cite{V}. On the other hand, the
\ne and \nm have a tiny mass difference in the model of
ref. \cite{MASI_pot3}. This mass difference can be chosen to lie
in the range where resonant \ne to \nm conversions provides
an explanation of solar \neu deficit \cite{MSW}. Due to this peculiar
hierarchical pattern, one can go even further, and regard the
rare R parity violating processes as a tool to probe the
physics underlying the solar \neu conversions in this model
\cite{RPMSW}. Indeed, the rates for such rare decays can be
used in order to discriminate between large and small mixing
angle MSW solutions to the solar \neu problem \cite{MSW}.
Typically, in the nonadiabatic region of small mixing one can have
larger rare decay branching ratios, as seen in Fig. 5
of ref. \cite{RPMSW}.

It is also possible to find manifestations of
R parity violation at the superhigh energies available
at hadron supercolliders such as LHC. Either SUSY
particles, such as gluinos, are pair produced and
in their cascade decays the LSP decays or, alternatively,
one violates R parity by singly producing the SUSY states.
An example of this situation has been discussed in
ref. \cite{RPLHC}. In this reference one has
studied the single production of weakly interacting
supersymmetric fermions (charginos and neutralinos)
via the Drell Yan mechanism, leading to possibly
detectable signatures. More work on this will
be desirable.

Another possible signal of the RPSUSY models based
on the simplest \21 gauge group is rare decays of
muons and taus. In this model the spontaneous
violation of R parity generates a physical Goldstone
boson, called majoron. Its existence is quite consistent
with the measurements of the invisible $Z$ decay width at LEP,
as it is a singlet under the \21 \gau symmetry.
In this model the lepton number is broken close
to the weak scale and can produce a new class of
lepton flavour violating decays, such as those
with single majoron emission in $\mu$ and $\tau$
decays. These would be "seen" as bumps in the final
lepton energy spectrum, at half of the parent lepton
mass in its rest frame.
\begin{table}
\begin{center}
\caption{Allowed branching ratios for rare decays
in the RPSUSY model. $\chi$ denotes the lightest
electrically charged SUSY fermion (chargino)
and $\chi^0$ is the lightest neutralino.}
\begin{displaymath}
\begin{array}{|c|cr|}
\hline
\mbox{channel} & \mbox{strength} & \mbox{} \\
\hline
Z \rightarrow \chi \tau &  \lsim 6 \times 10^{-5} & \\
Z \rightarrow \chi^0 \nt &  \lsim 10^{-4} & \\
\hline
\tau \rightarrow \mu + J &  \lsim 10^{-3} & \\
\tau \rightarrow e + J &  \lsim 10^{-4} & \\
\hline
\end{array}
\end{displaymath}
\end{center}
\end{table}
The allowed rates for single majoron emitting $\mu$
and $\tau$ decays have been determined in ref. \cite{NPBTAU}
and are also shown in table 3 to be compatible with present
experimental sensitivities \cite{PDG94}. Moreover, they are
ideally studied at a tau-charm factory \cite{TTTAU}.
This example also illustrates how the search for
rare decays can be a more sensitive probe of \neu
properties than the more direct searches for \neu
masses, and therefore complementary.

\section{Higgs Bosons.}
\hspace{0.5cm}

Another possible, albeit quite indirect, manifestation
of the properties of \neus and the lepton sector is in
the electroweak breaking sector. Many extensions of the
lepton sector seek to give masses to \neus through the
spontaneous violation of an ungauged U(1) lepton number
symmetry, thus implying the existence of a physical
Goldstone boson, called majoron \cite{CMP}. As already
mentioned above this is consistent with the measurements
of the invisible $Z$ decay width at LEP if the majoron
is (mostly) a singlet under the \21 \gau symmetry.

Although the original majoron proposal was made
in the framework of the minimal seesaw model, and
required the introduction of a relatively high
energy scale associated to the mass of the \rh
\neus \cite{CMP}, there are many attractive
theoretical alternatives where lepton number
is violated spontaneously at the weak scale or
lower. In this case although the majoron has very
tiny couplings to matter and the \gau bosons, it
can have significant couplings to the Higgs bosons.
As a result one has the possibility that the Higgs
boson may decay with a substantial branching ratio
into the invisible mode \cite{JoshipuraValle92}
\begin{equation}
h \rightarrow J\;+\;J
\label{JJ}
\end{equation}
where $J$ denotes the majoron. The presence of
this invisible decay channel can affect the
corresponding Higgs mass bounds in an important way.

The production and subsequent decay of a Higgs boson
which may decay visibly or invisibly involves three independent
parameters: its mass $M_H$, its coupling strength to the Z,
normalized by that of the \sm, $\epsilon^2$, and its
invisible decay branching ratio.
The LEP searches for various exotic channels can be used
in order to determine the regions in parameter space
that are already ruled out, as described in ref.
\cite{alfonso}. The exclusion contour in the plane
$\epsilon^2$ vs. $M_H$, was shown in Fig. 2 of
ref. \cite{moriond}.

Another mode of production of invisibly decaying
Higgs bosons is that in which a CP even Higgs boson
is produced at LEP in association with a massive
CP odd scalar \cite{HA}. This production mode is
present in all but the simplest majoron model
containing just one complex scalar singlet in addition
to the \sm Higgs doublet. Present limits on the
relevant parameters are given in ref. \cite{HA}.

Finally, the invisible decay of the Higgs boson may
also affect the strategies for searches at higher energies.
For example, the ranges of parameters that can be covered
by LEP2 searches for a total integrated luminosity of
500 pb$^{-1}$ and various centre-of-mass energies have
been given in Fig. 2 of the first paper in ref. \cite{alfonso}.
Similar analysis were made for the case of a high energy linear
$e^+ e^-$ collider (NLC) \cite{EE500}, as well as for the LHC
\cite{granada}.

\section{New Gauge Bosons.}
\hspace{0.5cm}

Superstring extensions of the standard model suggest the
existence of additional \gau bosons at the TeV scale and
this may affect the lepton sector and the interactions of
neutrinos. For example, an additional \ZP at
low energies would modify the couplings of leptons to the
Z and be thereby restricted by low energy neutral current
data \cite{CHARM}, as well as by the LEP precision data on
Z decays \cite{altarelli}. In string models the Higgs sector
is constrained in such a way that these limits are strongly
correlated with the top quark mass \cite{B259}. The recent
data from the CDF collaboration leads to contraints around
a TeV on the \ZP mass for various models of the string type
that fit in the $E_6$ \gau group. The limits are much weaker
in the case of unconstrained models.

\section*{Acknowledgements}
\hspace{0.5cm} This work has been supported by DGICYT under
Grant number PB92-0084.

\end{document}

%% file: matex.tex
\newcommand {\ignore}[1]{}

\newcommand{\bc}{\begin{center}}
\newcommand{\ec}{\end{center}}

\def\ifmath#1{\relax\ifmmode #1\else $#1$\fi}
\def\3quarter{{\textstyle{3 \over 4}}}

\def\lf{\leaders\hbox to 1em{\hss.\hss}\hfill}
\def\ZP{$Z^\prime$ }
\def\21{$SU(2) \ot U(1)$}
\def\321{$SU(3) \ot SU(2) \ot U(1)$}
\def\ne{\hbox{$\nu_e$ }}
\def\nm{\hbox{$\nu_\mu$ }}
\def\nt{\hbox{$\nu_\tau$ }}

\def\Nt{\hbox{$N_\tau$ }}

\def\O{\hbox{$\cal O$ }}

        \def\etc{\hbox{\it etc. }}
\def\eg{\hbox{\it e.g., }}        
\def\etal{\hbox{\it et al., }}

\def\rh{\hbox{right-handed }}

\def\gau{\hbox{gauge }}

\def\sm{\hbox{standard model }}

\def\neu{\hbox{neutrino }}
\def\sa{\hbox{such as }}

\def\neus{\hbox{neutrinos }}

\def\lsim{\raise0.3ex\hbox{$\;<$\kern-0.75em\raise-1.1ex\hbox{$\sim\;$}}}
\def\gsim{\raise0.3ex\hbox{$\;>$\kern-0.75em\raise-1.1ex\hbox{$\sim\;$}}}

\def\bel{\begin{letter}}
\def\eel{\end{letter}}
\def\beq{\begin{equation}}
\def\eeq{\end{equation}}
\def\bef{\begin{figure}}
\def\eef{\end{figure}}
\def\bet{\begin{table}}
\def\eet{\end{table}}
\def\bea{\begin{eqnarray}}
\def\ba{\begin{array}}
\def\ea{\end{array}}
\def\bi{\begin{itemize}}
\def\ei{\end{itemize}}
\def\ben{\begin{enumerate}}
\def\een{\end{enumerate}}

\def\ot{\otimes}

\def\eea{\end{eqnarray}}
\def\ib#1#2#3{           {\it ibid. }{\bf #1} (19#2) #3}

\def\nps#1#2#3{          {\it Nucl. Phys. B (Proc. Suppl.) }
                         {\bf #1} (19#2) #3}
\def\np#1#2#3{           {\it Nucl. Phys. }{\bf #1} (19#2) #3}
\def\pl#1#2#3{           {\it Phys. Lett. }{\bf #1} (19#2) #3}
\def\pr#1#2#3{           {\it Phys. Rev. }{\bf #1} (19#2) #3}
\def\prep#1#2#3{         {\it Phys. Rep. }{\bf #1} (19#2) #3}
\def\prl#1#2#3{          {\it Phys. Rev. Lett. }{\bf #1} (19#2) #3}

\def\n.c.#1#2#3{         {\it Nuovo Cim. }{\bf #1} (19#2) #3}
\def\r.n.c.#1#2#3{       {\it Riv. del Nuovo Cim. }{\bf #1} (19#2) #3}
\def\sjnp#1#2#3{         {\it Sov. J. Nucl. Phys. }{\bf #1} (19#2) #3}

\def\mpl#1#2#3{          {\it Mod. Phys. Lett. }{\bf #1} (19#2) #3}

\def\ppnp#1#2#3{           {\it Prog. Part. Nucl. Phys. }{\bf #1} (19#2) #3}

\relax